\documentclass[prb,twocolumn,showpacs]{revtex4}
\usepackage{graphicx} 
\newcommand{\etal}{{ \it et al. }}

\usepackage{amsmath,amsthm,amssymb,mathrsfs}
\usepackage{bm}

\begin{document}

\title{Magnetization reversal condition for a nanomagnet within a rotating magnetic field}

\author{Tomohiro Taniguchi}
 \affiliation{
 National Institute of Advanced Industrial Science and Technology (AIST), Spintronics Research Center, Tsukuba, Ibaraki 305-8568, Japan.
 }

 \date{\today} 
 \begin{abstract}
  {
    The reversal condition of magnetization in a nanomagnet under the effect of 
    rotating magnetic field generated by 
    a microwave is theoretically studied 
    based on the Landau-Lifshitz-Gilbert equation. 
    In a rotating frame, 
    the microwave produces a dc magnetic field pointing in the reversed direction, 
    which energetically stabilizes the reversed state. 
    We find that the microwave simultaneously produces a torque 
    preventing the reversal. 
    It is pointed out that this torque leads to a jump in the reversal field 
    with respect to the frequency. 
    We derive the equations determining the reversal fields in both the low- and high-frequency regions 
    from the energy balance equation. 
    The validities of the formulas are confirmed 
    by a comparison with the numerical simulation of the Landau-Lifshitz-Gilbert equation. 
  }
 \end{abstract}

 \maketitle


\section{Introduction}
\label{sec:Introduction}

Magnetization reversal in a single-domain ferromagnetic nanostructure 
is an important phenomenon for both fundamental physics and applications. 
The conventional method for reversing magnetization 
is to apply a direct magnetic field to a ferromagnet 
along the reversed direction, 
where the field magnitude $H$ should be larger than 
the anisotropy field (or coercivity) $H_{\rm K}$ 
to energetically stabilize the reversed state \cite{hubert98}. 
However, this method requires a large field $H$ 
anti-parallel to the magnetization, 
as well as large power consumption, for the reversal 
because ferromagnets with large $H_{\rm K}$ are used in practical applications 
to keep the high thermal stability $\Delta_{0}=MH_{\rm K}V/(2k_{\rm B}T)$, 
where $M$, $V$, and $T$ are the magnetization, the volume of the ferromagnet, and the temperature, 
respectively. 
Recently alternative methods, 
such as spin-torque-induced magnetization reversal \cite{slonczewski96,berger96,katine00,zhang02,kiselev03,kubota05,deac06,taniguchi08,taniguchi09} 
and microwave-assisted magnetization reversal (MAMR) 
\cite{bertotti01,bertotti01a,thirion03,denisov06,sun06,zhu08,bertotti09,bertotti09book,okamoto08,okamoto10,okamoto12,okamoto13,okamoto14,barros11,barros13,cai13}, 
have been proposed to reduce the reversal field magnitude. 
The optical magnetization reversal with circularly polarized light \cite{stanciu07,gevaux07} is 
another possibility, 
where the combination of the ultrafast heating and the magnetic field, 
both of which are generated by the circularly polarized laser, 
enables the ultrafast magnetization reversal without an external field. 


In MAMR, the microwave produces a circularly rotating magnetic field 
applied to the ferromagnet, 
in which the field direction lies in a plane 
perpendicular to the easy axis. 
A rotating frame is conventionally used to understand 
why the reversal field becomes smaller than $H_{\rm K}$ in MAMR \cite{bertotti09book,okamoto13}. 
In the rotating frame, 
the field acting on the magnetization is independent of time. 
The effect of the rotating field is converted to 
an additional dc magnetic field $(2\pi f/\gamma)$ pointing in the reversed direction \cite{bertotti01,denisov06,bertotti09book}, 
where $f$ and $\gamma$ are the frequency of the rotating field and the gyromagnetic ratio, respectively, 
i.e., the total dc magnetic field pointing in the reversed direction is $H+(2\pi f/\gamma)$. 
The additional field $(2\pi f/\gamma)$ energetically stabilizes the reversed state, 
and reduces the reversal field magnitude. 
In a low-frequency region, 
the reversal field linearly decreases as the frequency increases, 
which is qualitatively consistent with this conventional picture. 
However, both the experiments and the numerical simulations of 
the Landau-Lifshitz-Gilbert (LLG) equation have revealed that 
such a conventional picture cannot explain the dependence of the reversal field on the frequency 
in a high-frequency region \cite{zhu08,okamoto08,okamoto12,okamoto13}. 
In the high-frequency region, 
the reversal field slightly increases as the frequency increases; 
see, for example, Fig. \ref{fig:fig5} below. 
Moreover, the magnitude of the total dc magnetic field for the reversal, $H+(2\pi f/\gamma)$, 
in the high-frequency region is much larger than $H_{\rm K}$. 
This result seems like in contradiction with 
the Stoner-Wohlfarth theory \cite{hubert98}, 
in which the magnetization reversal should occur 
when $H+(2\pi f/\gamma)$ becomes slightly larger than $H_{\rm K}$ 
because only the reversed state is energetically stable. 


Okamoto \etal studied the dependence of the reversal field on the frequency of the rotating field 
for a Co/Pt nanodot with 120 nm diameter 
both experimentally and numerically based on the micromagnetic model \cite{okamoto12}. 
They found that the excitation of the spin wave, 
which arises from the difference in the local demagnetization field 
between the end and center of the dots, 
leads to a reduction of the reversal field. 
This result is of great advance in understanding the reversal mechanism of MAMR. 
However, it is still unclear why the reversal field jumps to a high value at a certain frequency. 
Moreover, the numerical simulations based on the macrospin (single domain) model also 
show the jump of the reversal field \cite{zhu08,okamoto12}, 
indicating that not only the excitation of the spin wave 
but also other mechanisms lead to this jump. 
A fabrication of the ferromagnet smaller than the exchange length (typically \cite{okamoto13}, on the order of 10 nm) 
is an unavoidable and indispensable challenge 
for both fundamental physics and practical applications. 
In such nanostructure, the magnetization dynamics is well described by the macrospin model. 
Therefore, it is important to clarify the magnetization reversal mechanism, 
such as the origin of the jump of the reversal field, 
using the macrospin model. 


The purposes of this paper are 
to explain why a large field is required to reverse the magnetization 
in the high-frequency region in MAMR 
and to derive equations that determine the reversal field 
in both the low- and high-frequency regions 
without the time-dependent solution of the macrospin LLG equation. 
It is pointed out that 
the rotating field not only energetically stabilizes the reversed state 
but also produces a torque acting on the magnetization. 
We find that this torque, 
whose strength is proportional to the frequency of the rotating field, 
prevents the reversal, 
causing the reversal field to become 
relatively large in the high-frequency region. 
The direction of this preventing torque is expressed 
by the triple vector product, 
analogous to the spin torque \cite{slonczewski96,berger96}. 
This fact motivates us 
to use the energy balance equation 
for the estimation of the reversal field of MAMR, 
which was recently pointed out to be useful 
for estimating the reversal current 
in the spin-torque reversal problem \cite{hillebrands06,newhall13,pinna13,taniguchi13,taniguchi13a,taniguchi13b,taniguchi14} 
but has been never applied to the MAMR problem. 
The equations determining the reversal field, 
Eqs. (\ref{eq:condition_low}) and (\ref{eq:condition_high}), are derived 
for both the low- and high-frequency regions 
from the energy balance equation. 
These formulas show that 
the reversal field in the low-frequency region 
converges to Eq. (\ref{eq:condition_low_zero_damping}) 
as the damping constant decreases, 
while the reversal field in the high-frequency region 
is independent of the damping constant. 
The boundary between the low- and high-frequency regions is 
also estimated from the energy balance equation. 
The validities of these formulas are quantitatively confirmed by comparison with 
the numerical simulation of the LLG equation. 


The paper is organized as follows. 
In Sec. \ref{sec:Landau-Lifshitz-Gilbert equation in rotating frame}, 
the energy balance equation is derived from the LLG equation 
in the rotating frame. 
The equations determining the reversal fields 
in the low- and high-frequency regions are derived 
in Secs \ref{sec:Reversal in low frequency region} and \ref{sec:Reversal in high frequency region}, respectively. 
The validities of these formulas over a wide frequency range are confirmed in  
Sec. \ref{sec:Comparison with numerical simulation} 
by comparison with the numerical simulation of the LLG equation. 
In Sec. \ref{sec:Comparison with other work},
the present result is compared with the previous work in Refs. \cite{bertotti01,bertotti01a,bertotti09,bertotti09book}. 
The conclusion appears in Sec. \ref{sec:Conclusion}. 


\section{Landau-Lifshitz-Gilbert equation in rotating frame}
\label{sec:Landau-Lifshitz-Gilbert equation in rotating frame}


\begin{figure}
\centerline{\includegraphics[width=1.0\columnwidth]{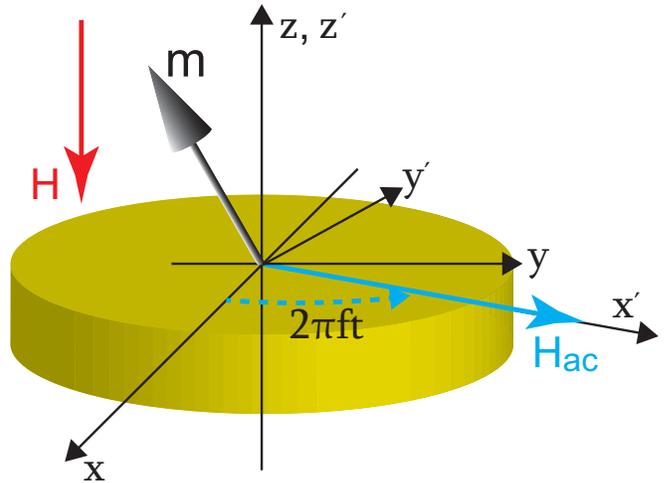}}\vspace{-3.0ex}
\caption{
         Schematic view of the system. 
         The unit vector pointing in the magnetization direction is denoted as $\mathbf{m}$. 
         The external magnetic field pointing in the negative $z$ direction 
         and the rotating field with the frequency $f$ are denoted as $H$ and $H_{\rm ac}$, respectively. 
         In the rotating frame, $x^{\prime}$ and $z^{\prime}$ axes are 
         parallel to the rotating field and the $z$ axis, respectively. 
         \vspace{-3ex}}
\label{fig:fig1}
\end{figure}


The system we consider is schematically shown in Fig. \ref{fig:fig1}, 
where the unit vector pointing in the magnetization direction is denoted as $\mathbf{m}$. 
The ferromagnet has a uniaxial easy axis 
with the anisotropy field $H_{\rm K}$ along the $z$ axis. 
Throughout this paper, 
the initial state is taken to be $\mathbf{m}=+\mathbf{e}_{z}$, 
although the following formulas are applicable to 
an arbitrary initial condition. 
The external field $H$ is applied to the negative $z$ direction. 
The rotating field with the magnitude $H_{\rm ac}$ and the frequency $f$ is applied 
in the $xy$ plane, 
where the $x$ axis is parallel to the rotating field 
at the initial time $t=0$. 
The magnetization dynamics 
under the effect of the magnetic field, 
$\mathbf{H}=H_{\rm ac} \cos (2\pi ft) \mathbf{e}_{x} 
           +H_{\rm ac} \sin (2\pi ft) \mathbf{e}_{y} 
           +(-H + H_{\rm K}m_{z})\mathbf{e}_{z}$ 
are described by 
the LLG equation \cite{landau35,landau80,gilbert04}, 
\begin{equation}
  \frac{d \mathbf{m}}{dt}
  =
  -\gamma
  \mathbf{m}
  \times
  \mathbf{H}
  +
  \alpha 
  \mathbf{m}
  \times
  \frac{d \mathbf{m}}{dt}, 
  \label{eq:LLG_orig}
\end{equation}
where the Gilbert damping constant is denoted as $\alpha$. 
Because the LLG equation conserves the magnitude of the magnetization, 
the magnetization dynamics are described by the trajectory 
on the surface of the unit sphere. 


It is convenient to use 
the rotating frame $x^{\prime}y^{\prime}z^{\prime}$, 
in which the $z^{\prime}$ axis is parallel to the $z$ axis, 
and $x^{\prime}$ axis is parallel to the rotating field $H_{\rm ac}$, 
as shown in Fig. \ref{fig:fig1}. 
We denote $\mathbf{m}$ in the rotating frame as $\mathbf{m}^{\prime}=(m_{x^{\prime}},m_{y^{\prime}},m_{z^{\prime}})$. 
It should be noted that the value of $m_{z}$ is invariant by this transformation. 
Because the value of $\alpha$ in the conventional ferromagnet is small \cite{oogane06}, 
higher order terms of $\alpha$ are neglected in the following; 
i.e., we use the approximation that $1+\alpha^{2}\simeq 1$. 
Then, the LLG equation in the rotating frame is given by 
\begin{equation}
\begin{split}
  \frac{d \mathbf{m}^{\prime}}{dt}
  =&
  -\gamma
  \mathbf{m}^{\prime}
  \times
  \bm{\mathcal{B}}
  -
  \alpha
  \gamma 
  \mathbf{m}^{\prime}
  \times 
  \left(
    \mathbf{m}^{\prime}
    \times
    \bm{\mathcal{B}}
  \right)
\\
  &+
  \alpha
  2\pi f 
  \mathbf{m}^{\prime}
  \times
  \left(
    \mathbf{e}_{z^{\prime}}
    \times
    \mathbf{m}^{\prime}
  \right), 
  \label{eq:LLG}
\end{split}
\end{equation}
where $\bm{\mathcal{B}}=H_{\rm ac} \mathbf{e}_{x^{\prime}} + [-H-(2\pi f/\gamma)+H_{\rm K}m_{z^{\prime}}] \mathbf{e}_{z^{\prime}}$ 
can be regarded as the magnetic field in the rotating frame. 
The transformation procedure from Eq. (\ref{eq:LLG_orig}) to Eq. (\ref{eq:LLG}) 
is shown in Appendix A. 


It can be understood from Eq. (\ref{eq:LLG}) that 
the rotating field plays two roles for the reversal. 
First, the magnetization dynamics can be regarded as 
a motion of a point particle 
in the potential $\mathscr{E}= -M \int d \mathbf{m} \cdot \bm{\mathcal{B}}$, 
\begin{equation}
  \mathscr{E}
  =
  -M H_{\rm ac}
  m_{x^{\prime}}
  +
  M
  \left(
    H 
    +
    \frac{2\pi f}{\gamma}
  \right)
  m_{z^{\prime}}
  -
  \frac{MH_{\rm K}}{2}
  m_{z^{\prime}}^{2}.
  \label{eq:potential} 
\end{equation}
The second term on the right-hand side of Eq. (\ref{eq:potential}) 
indicates that 
the rotating field produces the dc magnetic field $(2\pi f/\gamma)$ 
pointing in the negative $z^{\prime}$ direction, 
and energetically stabilizes the reversed state \cite{bertotti09,bertotti09book,okamoto13}. 
Second, the rotating field produces a torque 
proportional to the frequency $f$, 
which appears in the third term on the right-hand side of Eq. (\ref{eq:LLG}). 
The important point is that 
this torque points to the positive $z^{\prime}$ direction, 
and therefore, prevents the reversal. 
It should also be emphasized that 
the torque direction is expressed by the triple vector product, 
as is similar to the spin torque \cite{slonczewski96,berger96}. 
Therefore, in the following calculations, 
let us conventionally call this torque spin torque. 


It was shown in the spin-torque reversal problem \cite{hillebrands06,newhall13,pinna13,taniguchi13,taniguchi13a,taniguchi13b,taniguchi14} that 
the magnetization reversal condition can be derived 
from the energy balance equation 
between the work done by spin torque 
and the energy dissipation due to the damping. 
In the following sections, 
we apply this method to investigate 
the reversal field in MAMR. 
To this end, 
the derivative of $\mathscr{E}$ with respect to time 
on the constant energy curve is necessary. 
From Eq. (\ref{eq:LLG}), we find that 
\begin{equation}
\begin{split}
  \frac{d \mathscr{E}}{dt}
  =&
  -\alpha 
  2\pi f M 
  \left[
    \bm{\mathcal{B}}
    \cdot
    \mathbf{e}_{z^{\prime}}
    -
    \left(
      \mathbf{m}^{\prime}
      \cdot
      \mathbf{e}_{z^{\prime}}
    \right)
    \left(
      \mathbf{m}^{\prime}
      \cdot
      \bm{\mathcal{B}}
    \right)
  \right]
\\
  &-
  \alpha 
  \gamma 
  M
  \left[
    \bm{\mathcal{B}}^{2}
    -
    \left(
      \mathbf{m}^{\prime}
      \cdot
      \bm{\mathcal{B}}
    \right)^{2}
  \right]. 
  \label{eq:dEdt}
\end{split}
\end{equation}
The integral of Eq. (\ref{eq:dEdt}) over a precession period of the magnetization 
on the constant energy curve of $\mathscr{E}$ is 
$\oint dt (d \mathscr{E}/dt)=\mathscr{W}_{\rm s}+\mathscr{W}_{\alpha}$, 
where 
\begin{equation}
  \mathscr{W}_{\rm s}
  =
  -\alpha 
  2\pi f M 
  \oint dt 
  \left[
    \bm{\mathcal{B}}
    \cdot
    \mathbf{e}_{z^{\prime}}
    -
    \left(
      \mathbf{m}^{\prime}
      \cdot
      \mathbf{e}_{z^{\prime}}
    \right)
    \left(
      \mathbf{m}^{\prime}
      \cdot
      \bm{\mathcal{B}}
    \right)
  \right],
  \label{eq:W_s}
\end{equation}
\begin{equation}
  \mathscr{W}_{\alpha}
  =
  -\alpha
  \gamma 
  M 
  \oint dt 
  \left[
    \bm{\mathcal{B}}^{2}
    -
    \left(
      \mathbf{m}^{\prime}
      \cdot
      \bm{\mathcal{B}}
    \right)^{2}
  \right],
  \label{eq:W_alpha}
\end{equation}
are the work done by spin torque 
and the energy dissipation due to the damping 
during the precession, respectively \cite{perko91}. 
While $\mathscr{W}_{\rm s}$ can be both positive and negative 
depending on the field and the frequency, 
$\mathscr{W}_{\alpha}$ is always negative. 
The calculation procedures of Eqs. (\ref{eq:W_s}) and (\ref{eq:W_alpha}) 
without the time-dependent solution of Eq. (\ref{eq:LLG}) 
are shown in Appendix B. 
The damping constant $\alpha$ is assumed to be scalar 
in the above formulation. 
On the other hand, Safonov studied the magnetization relaxation near equilibrium 
with a tensor damping \cite{safonov02}. 
The presence of the tensor damping was also suggested 
in the spin torque problem \cite{zhang09}. 
In Appendix C, we discuss how the formulas derived in the following sections are modified 
when the scalar damping $\alpha$ is replaced by the tensor damping. 


\section{Reversal in low-frequency region}
\label{sec:Reversal in low frequency region}

\begin{figure}
\centerline{\includegraphics[width=1.0\columnwidth]{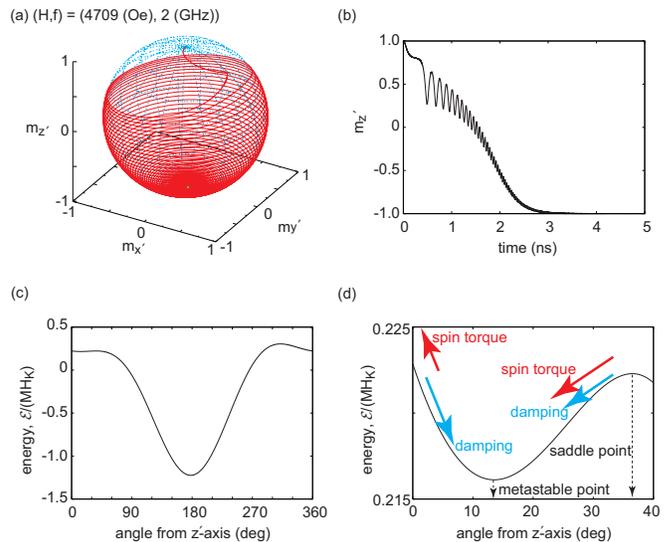}}\vspace{-3.0ex}
\caption{
         (a) Schematic view of the magnetization dynamics in the rotating frame 
             at the reversal field $H=4.709$ kOe, 
             where the frequency of the rotating field is $f=2.0$ GHz. 
             The trajectory is described on the unit sphere. 
         (b) The time evolution of $m_{z^{\prime}}$. 
         (c) Schematic view of the potential $\mathscr{E}$ (normalized by $MH_{\rm K})$ 
             in the $x^{\prime}z^{\prime}$ plane. 
             The horizontal axis represents the angle $\theta$ of the magnetization from the $z^{\prime}$ axis. 
         (d) The enlarged view of panel (c) around $[\theta_{\rm initial},\theta_{\rm saddle}]$. 
             The directions of the damping and spin torque are indicated by arrows. 
         \vspace{-3ex}}
\label{fig:fig2}
\end{figure}

In this section, 
we study the reversal field in the low-frequency region. 
Let us first show in Fig. \ref{fig:fig2} (a) 
the trajectory of typical magnetization dynamics 
in the low-frequency region 
obtained by numerically solving Eq. (\ref{eq:LLG}). 
The time evolution of $m_{z^{\prime}}$ is shown in Fig. \ref{fig:fig2} (b). 
The values of the parameters are 
$M=1000$ emu/c.c., 
$H_{\rm K}=7.5$ kOe, 
$H_{\rm ac}=450$ Oe, 
$\gamma=1.764\times 10^{7}$ rad/(Oe$\cdot$s), 
$f=2.0$ GHz, 
and $\alpha=0.01$, 
which are typical values used in the experiments and the numerical simulations \cite{denisov06,zhu08,okamoto08,okamoto12,okamoto13}. 
We judged that the magnetization is reversed 
when the condition $m_{z^{\prime}}<-0.9$ is satisfied.  
The minimum field satisfying this condition is $H=4.709$ kOe. 
Starting from the initial state $\mathbf{m}^{\prime}=+\mathbf{e}_{z^{\prime}}$, 
the magnetization precesses around an axis 
lying in the positive $x^{\prime}z^{\prime}$ plane. 
After a half period of precession, 
the magnetization changes the precession direction, 
and falls into the reversed state. 


Next, let us analytically derive the equation determining the reversal field. 
Figure \ref{fig:fig2} (c) shows the map of the potential $\mathscr{E}$ 
in the $x^{\prime}z^{\prime}$ plane, 
where the horizontal axis is the angle $\theta$ of the magnetization from the $z^{\prime}$ axis. 
The potential with $H=4.709$ kOe has metastable, 
saddle, and stable points at 
$\theta_{\rm metastable}=14^{\circ}$, 
$\theta_{\rm saddle}=36^{\circ}$, 
and $\theta_{\rm stable}=178^{\circ}$, respectively. 
The directions of the damping and the spin torque 
between the initial state ($\theta_{\rm initial}=0^{\circ}$) and the saddle point 
in the potential $\mathscr{E}$ 
are indicated by 
the arrows in Fig. \ref{fig:fig2} (d). 
The spin torque supplies the energy to the ferromagnet 
when the magnetization is in $\theta_{\rm initial} \le \theta \le \theta_{\rm metastable}$ 
because it is anti-parallel to the damping, 
while the spin torque dissipates the energy when the magnetization is in $\theta_{\rm metastable} \le \theta \le \theta_{\rm saddle}$ 
because it is parallel to the damping. 
The function $\mathscr{W}_{\rm s}$ in $\theta_{\rm initial} \le \theta \le \theta_{\rm saddle}$ is negative 
($\mathscr{W}_{\rm s}<0$) 
because the spin torque magnitude ($\propto \sin\theta$) increases 
as the angle $\theta(<90^{\circ})$ increases. 
Therefore, the spin torque totally dissipates the energy during the precession. 
The magnetization reversal occurs 
when the magnitude of 
the energy $\Delta \mathscr{E}=\int dt (d \mathscr{E}/dt)$ dissipated 
during the dynamics from $\theta_{\rm initial}$ to $\theta_{\rm saddle}$ 
is smaller than the energy difference 
between the initial state and the saddle point, 
$\mathscr{E}_{\rm initial}-\mathscr{E}_{\rm saddle}$, 
i.e., the reversal condition is 
\begin{equation}
  \mathscr{E}_{\rm initial}
  -
  \mathscr{E}_{\rm saddle}
  \ge 
  -\Delta \mathscr{E}. 
  \label{eq:reversal_condition_low}
\end{equation}
Strictly speaking, 
the time-dependent solution of $\mathbf{m}^{\prime}$ is necessary to calculate $\Delta \mathscr{E}$. 
However, in the low-frequency region, 
because the difference between $\mathscr{E}_{\rm initial}$ and $\mathscr{E}_{\rm saddle}$ is small, 
$\Delta \mathscr{E}$ can be approximated to 
$[\mathscr{W}_{\rm s}(\mathscr{E}_{\rm saddle}) + \mathscr{W}_{\alpha}(\mathscr{E}_{\rm saddle})]/2$. 
The numerical factor $1/2$ appears 
because the reversal occurs after the half period of the precession. 
Therefore, the reversal field $H_{\rm reversal}$ can be defined as 
the field $H$ satisfying the condition 
\begin{equation}
  \mathscr{E}_{\rm initial}
  -
  \mathscr{E}_{\rm saddle}
  =
  -\frac{1}{2}
  \left[
    \mathscr{W}_{\rm s}(\mathscr{E}_{\rm saddle})
    +
    \mathscr{W}_{\alpha}(\mathscr{E}_{\rm saddle})
  \right].
  \label{eq:condition_low}
\end{equation}
Equation (\ref{eq:condition_low}) is the main result in this section. 
The reversal field estimated from Eq. (\ref{eq:condition_low}) is 4.708 kOe, 
which is almost identical to that (4.709 kOe) obtained 
from the numerical solution of the LLG equation (\ref{eq:LLG}). 
It should be noted that 
the approximation $\Delta \mathscr{E} \simeq [\mathscr{W}_{\rm s}(\mathscr{E}_{\rm saddle}) + \mathscr{W}_{\alpha}(\mathscr{E}_{\rm saddle})]/2$ 
works well for small $\alpha$ 
because the magnetization dynamics occurs almost on the constant energy curve 
when $\alpha \ll 1$. 
In the zero-damping limit, 
the reversal field is estimated from the condition $\mathscr{E}_{\rm initial}-\mathscr{E}_{\rm saddle}=0$, 
and is given by 
\begin{equation}
  H_{\rm reversal}
  =
  \frac{H_{\rm K}}{2}
  \sin^{2}\theta_{\rm saddle}
  \left(
    \frac{1}{\cos\theta_{\rm saddle}}
    -
    1
  \right)^{-1}
  -
  \frac{2\pi f}{\gamma},
  \label{eq:condition_low_zero_damping}
\end{equation}
where $\theta_{\rm saddle}$ depends on $H_{\rm reversal}$ 
through the condition 
$H_{\rm ac}\cos\theta_{\rm saddle}+[H_{\rm reversal}+(2\pi f/\gamma)]\sin\theta_{\rm saddle}-H_{\rm K} \sin\theta_{\rm saddle}\cos\theta_{\rm saddle}=0$. 


\begin{figure}
\centerline{\includegraphics[width=1.0\columnwidth]{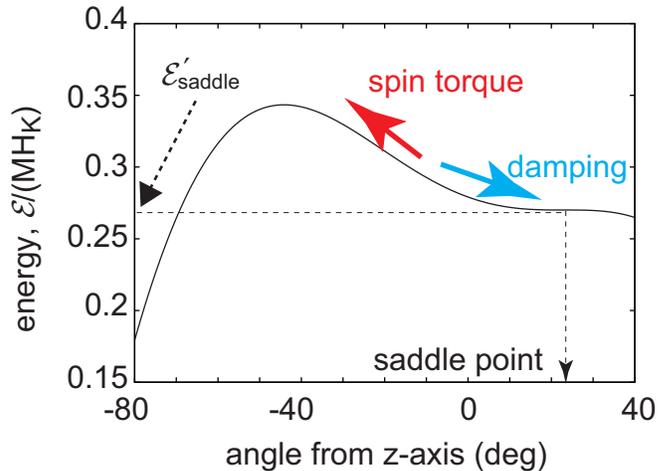}}\vspace{-3.0ex}
\caption{
         Schematic view of the potential $\mathscr{E}^{\prime}=\mathscr{E}(H=H^{\prime})$ 
         with $(H^{\prime},f)=(3.671 {\rm \ kOe},6.1 {\rm \ GHz})$. 
         The arrows indicate the directions of the spin torque and the damping. 
         \vspace{-3ex}}
\label{fig:fig3}
\end{figure}



Let us also elaborate on the frequency range 
in which Eq. (\ref{eq:condition_low}) obtains 
good agreement with the numerical solution of the LLG equation (\ref{eq:LLG}). 
At a certain field magnitude $H^{\prime}$, 
which is larger than $H_{\rm reversal}$ estimated by Eq. (\ref{eq:condition_low}), 
the metastable state disappears, 
and the potential has only one stable point and a saddle point. 
We denote the potential $\mathscr{E}$ with $H=H^{\prime}$ as $\mathscr{E}^{\prime}=\mathscr{E}(H=H^{\prime})$, 
which satisfies 
$\partial \mathscr{E}^{\prime}/\partial\theta=\partial^{2}\mathscr{E}^{\prime}/\partial\theta^{2}=0$ 
at the saddle point $\theta_{\rm saddle}^{\prime}$. 
According to the conventional Stoner-Wohlfarth theory \cite{hubert98}, 
the magnetization reversal should occur 
because the potential has only one minimum. 
However, in the present case, 
the work done by spin torque, $\mathscr{W}_{\rm s}(\mathscr{E}_{\rm saddle}^{\prime})$,
on the constant energy curve of 
$\mathscr{E}^{\prime}(\theta_{\rm saddle}^{\prime})=\mathscr{E}_{\rm saddle}^{\prime}$, 
is positive 
because the direction of the spin torque is always opposite 
that of the damping, as shown in Fig. \ref{fig:fig3}. 
Then, the condition 
\begin{equation}
  \mathscr{W}_{\rm s}(\mathscr{E}_{\rm saddle}^{\prime})
  +
  \mathscr{W}_{\alpha}(\mathscr{E}_{\rm saddle}^{\prime}) 
  \le 
  0, 
  \label{eq:condition_boundary}
\end{equation}
should also be satisfied to reverse the magnetization: 
if Eq. (\ref{eq:condition_boundary}) is not satisfied, 
the spin torque preventing the reversal overcomes the damping, 
and thus the magnetization cannot reverse its direction. 
It is found that Eq. (\ref{eq:condition_boundary}) is satisfied for $f<6.2$ GHz 
for the above parameters. 
Therefore, we define the low-frequency region 
in which Eq. (\ref{eq:condition_low}) is valid as $f<6.2$ GHz. 
Because the reversal field discontinuously becomes large above this frequency, 
we call this frequency the jump frequency. 
It should be emphasized that 
the jump frequency is independent of $\alpha (\ll 1)$. 



\section{Reversal in high-frequency region}
\label{sec:Reversal in high frequency region}

\begin{figure}
\centerline{\includegraphics[width=1.0\columnwidth]{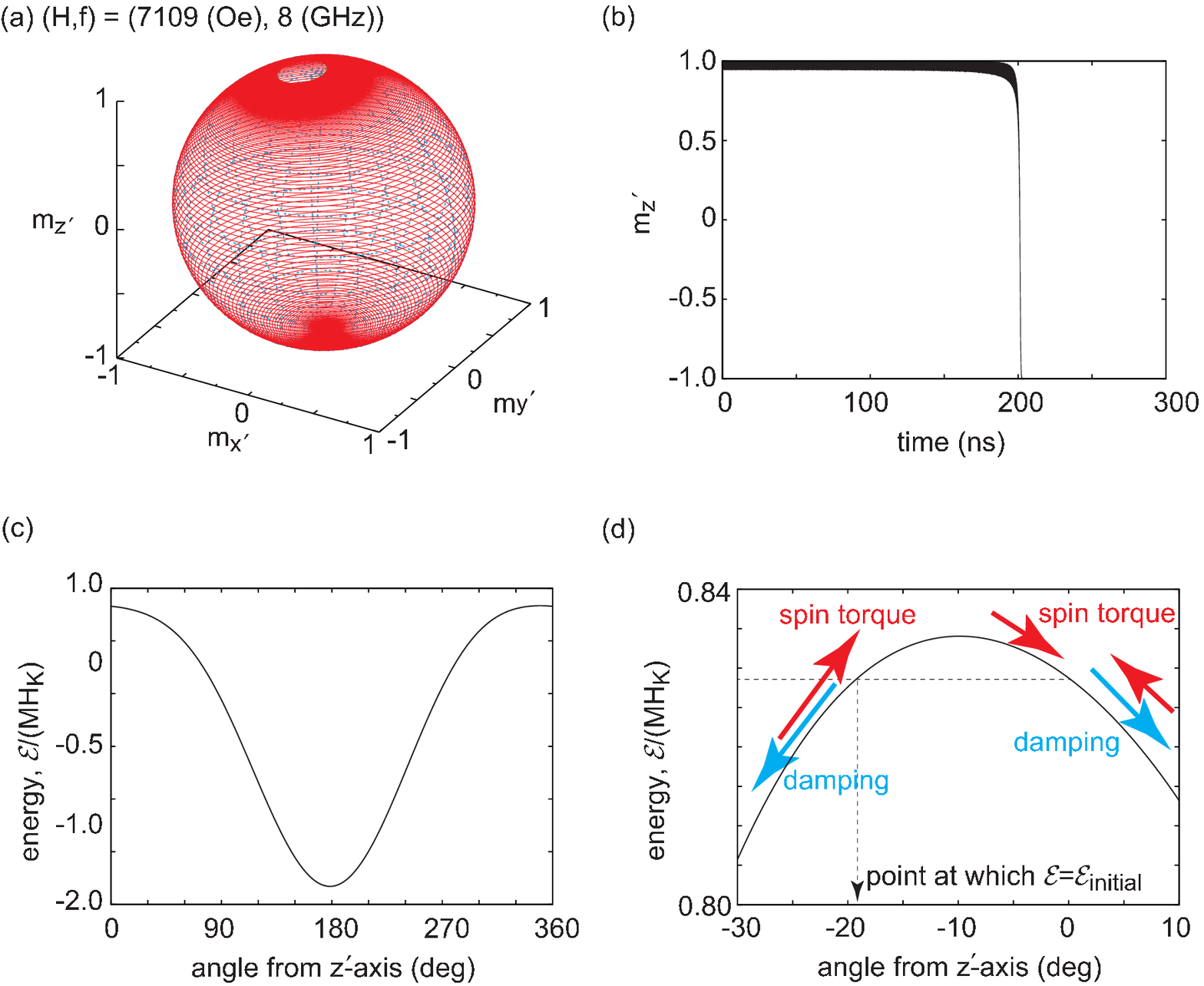}}\vspace{-3.0ex}
\caption{
         (a) Schematic view of the magnetization dynamics in the rotating frame 
             at the reversal field $H=7.109$ kOe, 
             where the frequency of the rotating field is $f=8.0$ GHz. 
             The trajectory is described on the unit sphere. 
         (b) The time evolution of $m_{z^{\prime}}$. 
         (c) Schematic view of the potential $\mathscr{E}$ (normalized by $MH_{\rm K})$ 
             in the $x^{\prime}z^{\prime}$ plane. 
             The horizontal axis represents the angle $\theta$ of the magnetization from the $z^{\prime}$ axis. 
         (d) The enlarged view of $\mathscr{E}$ around $-30^{\circ} \le \theta \le 10^{\circ}$. 
             The directions of the damping and spin torque are indicated by arrows. 
         \vspace{-3ex}}
\label{fig:fig4}
\end{figure}

In this section, 
we study the magnetization reversal in the high-frequency region. 
As mentioned in the previous section, 
for $f \ge 6.2$ GHz, 
the spin torque preventing the reversal becomes sufficiently large. 
Then, a large field is required 
to reverse the magnetization. 
Figure \ref{fig:fig4} (a) shows 
the trajectory of typical magnetization dynamics 
in the high-frequency region 
obtained by numerically solving Eq. (\ref{eq:LLG}). 
The time evolution of $m_{z^{\prime}}$ is shown in Fig. \ref{fig:fig4} (b). 
The values of the parameters are those used in Sec. \ref{sec:Reversal in low frequency region} 
except $f=8.0$ GHz. 
The minimum field satisfying the condition $m_{z^{\prime}}<-0.9$ is $H=7.109$ kOe. 
Starting from the initial state, 
the magnetization precesses on the constant energy curves 
near the $z^{\prime}$ axis many times. 
The precession amplitude slightly increases with time, 
and finally the magnetization reverses to $\mathbf{m}\simeq -\mathbf{e}_{z^{\prime}}$. 
The reversal trajectory covers almost all of the unit sphere, 
as shown in Fig. \ref{fig:fig4} (a). 


Figure \ref{fig:fig4} (c) shows the potential $\mathscr{E}$ in the $x^{\prime}z^{\prime}$-plane 
at the reversal field. 
According to the Stoner-Wohlfarth condition \cite{hubert98} 
$H_{\rm K}^{2/3}=[H+(2\pi f/\gamma)]^{2/3}+H_{\rm ac}^{2/3}$, 
a field $H$ larger than $3.0$ kOe is enough to reverse the magnetization, 
above which the potential $\mathscr{E}$ has only one minimum. 
Nevertheless, 
a large field ($\ge 7.109$ kOe) compared with the Stoner-Wohlfarth condition 
is required for the reversal in the present case 
because the spin torque prevents the reversal. 
Figure \ref{fig:fig4} (d) shows the enlarged view of $\mathscr{E}$ near the initial state, 
where the arrows indicate the directions of the damping and the spin torque. 
The maximum of $\mathscr{E}$ is located at $\theta_{\rm maximum}=-10^{\circ}$ 
while the angle satisfying $\mathscr{E}(\theta)=\mathscr{E}(\theta_{\rm initial})$ is located at $\theta=-19^{\circ}$. 
For a reason similar to that discussed in Sec. \ref{sec:Reversal in low frequency region}, 
the function $\mathscr{W}_{\rm s}(\mathscr{E}_{\rm initial})$ is positive. 
When $\oint dt (d \mathscr{E}/dt)=\mathscr{W}_{\rm s}(\mathscr{E}_{\rm initial})+\mathscr{W}_{\alpha}(\mathscr{E}_{\rm initial})<0$, 
the magnetization loses the energy, and falls into the reversed state. 
On the other hand, 
when $\mathscr{W}_{\rm s}(\mathscr{E}_{\rm initial})+\mathscr{W}_{\alpha}(\mathscr{E}_{\rm initial})>0$, 
the magnetization climbs $\mathscr{E}$ from $\theta_{\rm initial}$ to $\theta_{\rm maximum}$. 
Therefore, the reversal field can be defined as the field satisfying the condition 
\begin{equation}
  \mathscr{W}_{\rm s}(\mathscr{E}_{\rm initial})
  +
  \mathscr{W}_{\alpha}(\mathscr{E}_{\rm initial})
  =
  0.
  \label{eq:condition_high}
\end{equation}
Equation (\ref{eq:condition_high}) is the main result in this section. 
The reversal field estimated from Eq. (\ref{eq:condition_high}) is 7.109 kOe, 
which is identical to that obtained from the numerical solution of the LLG equation (\ref{eq:LLG}). 
Another important conclusion from Eq. (\ref{eq:condition_high}) is that 
the reversal field is independent of $\alpha$ 
because both $\mathscr{W}_{\rm s}$ and $\mathscr{W}_{\alpha}$ are proportional to $\alpha$. 
The validity of this conclusion is investigated in the next section. 


\section{Comparison with numerical simulation}
\label{sec:Comparison with numerical simulation}

\begin{figure}
\centerline{\includegraphics[width=1.0\columnwidth]{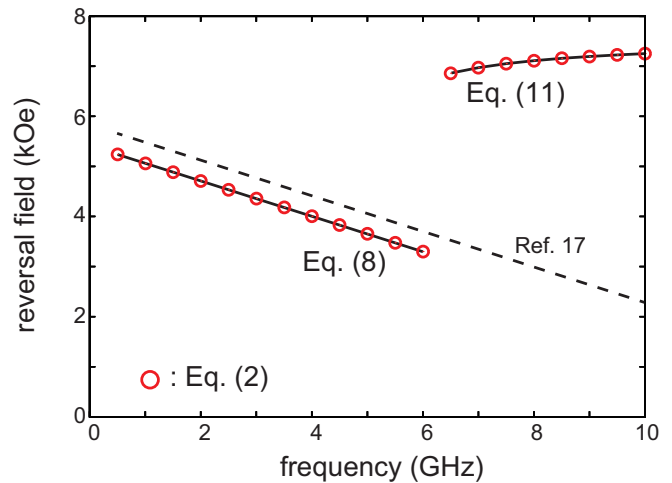}}\vspace{-3.0ex}
\caption{
         The dependence of the reversal field on the frequency of the rotating field. 
         The circles are obtained from Eq. (\ref{eq:LLG}) 
         while the solid lines are obtained from Eqs. (\ref{eq:condition_low}) and (\ref{eq:condition_high}). 
         The value of $\alpha$ is $0.01$.
         The dashed line is the reversal field estimated from Ref. \cite{bertotti09}, 
         and is discussed in Sec. \ref{sec:Comparison with other work}. 
         \vspace{-3ex}}
\label{fig:fig5}
\end{figure}

In this section, 
we confirm the validities of Eqs. (\ref{eq:condition_low}) and (\ref{eq:condition_high}) 
over a wide range of the frequency $f$. 
The circles in Fig. \ref{fig:fig5} show 
the reversal field estimated numerically solving the LLG equation (\ref{eq:LLG}), 
where the frequency range is $0 < f \le 10$ GHz. 
The Gilbert damping constant is $0.01$. 
The reversal field magnitude linearly decreases 
as the frequency increases for $f \lesssim 6$ GHz. 
Above $f \gtrsim 6$ GHz, 
the reversal field jumps to a high value 
at which $H_{\rm reversal}+(2\pi f/\gamma)>H_{\rm K}$, 
and slightly increases as the frequency increases. 
The reversal fields obtained from Eqs. (\ref{eq:condition_low}) and (\ref{eq:condition_high}) 
are also shown in Fig. \ref{fig:fig5} by the solid lines. 
As mentioned in Sec. \ref{sec:Reversal in low frequency region}, 
Eq. (\ref{eq:condition_low}) is valid for $f<6.2$ GHz. 
Therefore, we use Eq. (\ref{eq:condition_low}) for $f<6.2$ GHz 
while Eq. (\ref{eq:condition_high}) is used for $f \ge 6.2$ GHz. 
Equations (\ref{eq:condition_low}) and (\ref{eq:condition_high}) show good agreement 
with the numerical solution of the LLG equation (\ref{eq:LLG}), 
indicating the validities of these formulas. 


\begin{figure}
\centerline{\includegraphics[width=0.7\columnwidth]{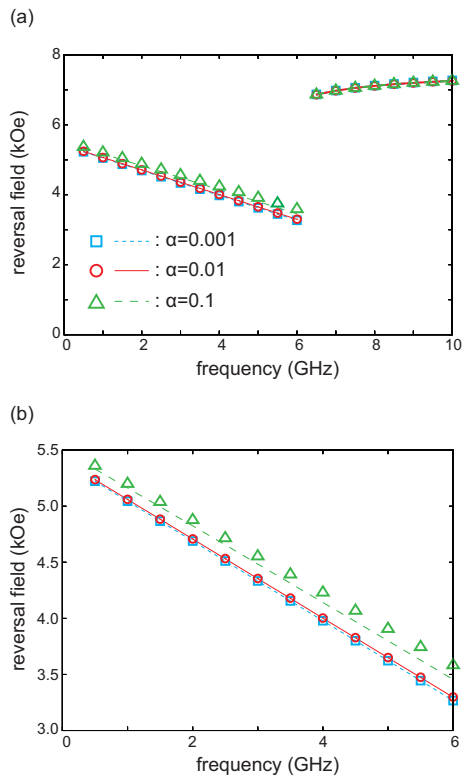}}\vspace{-3.0ex}
\caption{
         (a) The dependences of the reversal fields on the frequency for $\alpha=0.001$, $0.01$, and $0.1$. 
         The symbols (square, circle, and triangle) are obtained from Eq. (\ref{eq:LLG}) 
         while the lines (dotted, solid, and dashed) are obtained from Eqs. (\ref{eq:condition_low}) and (\ref{eq:condition_high}). 
         (b) The enlarged view of panel (a) in the low-frequency region. 
         \vspace{-3ex}}
\label{fig:fig6}
\end{figure}


Figure \ref{fig:fig6} (a) shows 
the dependences of the reversal fields on the frequency 
for $\alpha=0.001$, $0.01$, and $0.1$, 
where the numerical solution of the LLG equation (\ref{eq:LLG}) is represented by the symbols 
(square, circle, and triangle, respectively), 
while the reversal fields obtained from Eqs. (\ref{eq:condition_low}) and (\ref{eq:condition_high}) are 
represented by the lines (dotted, solid, and dashed, respectively). 
The frequency ($\simeq 6$ GHz) at which the reversal field of the LLG equation (\ref{eq:LLG}) 
jumps to a high value is independent of $\alpha$, 
which is consistent with the discussion in Sec. \ref{sec:Reversal in low frequency region}. 
The enlarged view in the low-frequency region is shown in Fig. \ref{fig:fig6} (b). 
In the low-frequency region, 
the difference between the solutions of the LLG equation (\ref{eq:LLG}) 
and the energy balance equation (\ref{eq:condition_low}) becomes small 
as $\alpha$ decreases, 
because the approximation $\Delta\mathscr{E} \simeq [\mathscr{W}_{\rm s}(\mathscr{E}_{\rm saddle}) + \mathscr{W}_{\alpha}(\mathscr{E}_{\rm saddle})]/2$ 
used in the derivation of Eq. (\ref{eq:condition_low}) is valid 
for a sufficiently small $\alpha$. 
Also, the reversal field becomes independent of $\alpha$ 
with decreasing $\alpha$, 
which is consistent with Eq. (\ref{eq:condition_low_zero_damping}). 
In the high-frequency region, 
the solution of the LLG equation (\ref{eq:LLG}) is also independent of $\alpha$, 
which is consistent with Eq. (\ref{eq:condition_high}). 
These results also imply the validity of Eqs. (\ref{eq:condition_low}) and (\ref{eq:condition_high}). 



\begin{figure}
\centerline{\includegraphics[width=1.0\columnwidth]{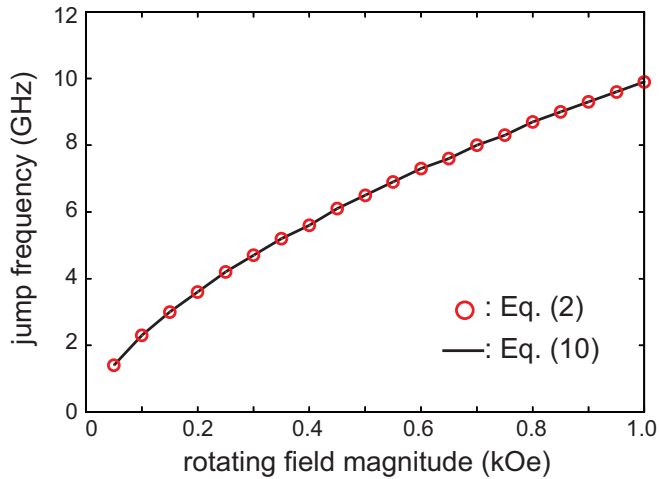}}\vspace{-3.0ex}
\caption{
         The dependence of the jump frequency, which is the boundary between the low- and high-frequency regions, 
         on the rotating field magnitude $H_{\rm ac}$. 
         The circles are obtained from Eq. (\ref{eq:LLG}) 
         while the solid line is obtained from Eq. (\ref{eq:condition_boundary}). 
         \vspace{-3ex}}
\label{fig:fig7}
\end{figure}


As we end this section, 
we study the relation between 
the rotating field magnitude $H_{\rm ac}$ 
and the jump frequency, i.e., the frequency defining the boundary 
between the low- and high-frequency regions. 
Figure \ref{fig:fig7} shows the dependences of the jump frequency on $H_{\rm ac}$ 
obtained from Eq. (\ref{eq:condition_boundary}) (solid line) 
and the numerical solution of the LLG equation (circle). 
As shown, the jump frequency monotonically increases with increasing $H_{\rm ac}$. 
Although the clarification of the relation between the jump frequency and the other parameters such as $H_{\rm ac}$ is desirable, 
it is difficult to analytically solve Eq. (\ref{eq:condition_boundary}) with respect to the jump frequency. 
We consider that the jump frequency does not necessarily relate to, for example, 
the ferromagnetic resonance (FMR) frequency 
because the jump frequency is determined by the energy balance of the magnetization at the saddle point of the potential $\mathscr{E}$ 
while the FMR frequency is the frequency of the harmonic oscillation around the stable point. 
However, a further investigation on the jump frequency is beyond the scope of this paper. 


\section{Comparison with other work}
\label{sec:Comparison with other work}

In this section, we compare the above result 
with the previous work of Bertotti \etal \cite{bertotti09,bertotti09book}. 
They expanded the LLG equation around its steady-state point, $(\theta,\phi)$, 
where $\theta$ and $\phi$ are the zenith and azimuth angles characterizing the magnetization direction, 
and satisfy the following equations: 
\begin{equation}
  \gamma 
  H_{\rm ac}
  \sin\phi
  -
  \alpha 
  2\pi f
  \sin\theta
  =
  0,
  \label{eq:condition_Bertotti_1}
\end{equation} 
\begin{equation}
  \left(
    \gamma 
    H 
    +
    2\pi f
  \right)
  \sin\theta
  -
  \gamma 
  H_{\rm K}
  \sin\theta
  \cos\theta
  +
  \gamma
  H_{\rm ac}
  \cos\theta
  \cos\phi
  =
  0.
  \label{eq:condition_Bertotti_2}
\end{equation}
Small deviation of the magnetization, $(\Delta\theta,\Delta\phi)$, from the steady points 
satisfy $d \Delta\theta/dt = \mathsf{A}_{1,1}\Delta\theta + \mathsf{A}_{1,2}\Delta\phi$ 
and $d \Delta\phi/dt = \mathsf{A}_{2,1}\Delta\theta + \mathsf{A}_{2,2} \Delta\phi$, 
where components of a $2 \times 2$ matrix $\mathsf{A}$ is obtained from the LLG equation. 
The trace and determinant of $\mathsf{A}$ are given by 
\begin{equation}
  {\rm Tr}[\mathsf{A}]
  =
  -\frac{2 \alpha}{1+\alpha^{2}}
  \left[
    v
    -
    \frac{\sin^{2}\theta}{2}
    \gamma
    H_{\rm K}
    +
    2\pi f 
    \cos\theta
  \right],
  \label{eq:trace_Bertotti}
\end{equation}
\begin{equation}
  {\rm det}[\mathsf{A}]
  =
  \frac{v^{2}-\gamma H_{\rm K} v \sin^{2}\theta + (\alpha 2\pi f \cos\theta)^{2}}{1+\alpha^{2}},
  \label{eq:determinant_Bertotti}
\end{equation}
where $v=\alpha 2\pi f \cot\phi$. 
According to Ref. \cite{bertotti09}, 
the reversal field is estimated 
from Eqs. (\ref{eq:condition_Bertotti_1}), (\ref{eq:condition_Bertotti_2}) 
and (\ref{eq:determinant_Bertotti}) with the condition ${\rm det}[\mathsf{A}]=0$. 
The dashed line in Fig. \ref{fig:fig5} is the reversal field estimated from this method. 
As shown, the method of Bertotti \etal reveals larger reversal field in our simulation. 
The difference between our and their results arises from the following reason. 
In our analytical and numerical calculations, 
both the microwave and external field are applied from $t=0$ with the constant magnitudes. 
The initial state of the magnetization, $m_{z^{\prime}}=1$, locates above the stable or saddle point, 
as shown in Fig. \ref{fig:fig2} (d). 
On the other hand, Bertotti \etal considered the instability of the magnetization 
around the steady point corresponding to the stable or saddle point. 
In this case, a relatively large energy compared with our situation is required to overcome the potential barrier and reverse the magnetization direction. 
Therefore, the reversal field estimated from the method of Bertotti \etal becomes larger than our estimation. 


We emphasize that both the results of Bertotti \etal and our method are useful to estimate the reversal field. 
For example, numerical simulation of the LLG equation \cite{okamoto10} 
showed good agreement with the theory of Bertotti \etal \cite{bertotti01,bertotti01a,bertotti09,bertotti09book}. 
In Ref. \cite{okamoto10}, 
the magnitude of the dc magnetic field is linearly increased with time until it reaches a certain value. 
In this case, the magnetization first relaxes to a stable point of the potential, 
and after that the magnetization reverses its direction 
when the saturated value of the dc magnetic field is larger than 
the reversal field estimated by the method of Bertotti \etal
On the other hand, our approach is applicable when 
the magnitude of the dc magnetic field is fixed from $t=0$, as mentioned above. 
To clarify the applicability of the theory of Bertotti \etal more precisely, 
an estimation of relaxation time from the initial state to the steady point, 
which should be shorter than the time to saturate the dc magnetic field magnitude, will be necessary. 


\section{Conclusion}
\label{sec:Conclusion}

In conclusion, 
we studied the dependence of the reversal field 
in microwave-assisted magnetization reversal 
on the frequency of the rotating field theoretically. 
The microwave produced a dc magnetic field pointing in the reversed direction, 
which energetically stabilized the reversed state. 
The microwave simultaneously produced a torque 
proportional to the frequency of the rotating field. 
Because this torque prevented the reversal, 
a large field was required to reverse the magnetization 
in the high-frequency region. 
The equations determining the reversal fields 
in both the low- and high-frequency regions were derived 
from the energy balance equation. 
The formulas showed that 
the reversal field in the low-frequency region became 
converged to Eq. (\ref{eq:condition_low_zero_damping}) 
as the damping constant decreased, 
while the reversal field in the high-frequency region 
was independent of the damping constant. 
The boundary between the low- and high-frequency regions, 
which was independent of the damping constant, was 
also estimated from the energy balance equation. 
The comparison with the numerical solution of the Landau-Lifshitz-Gilbert equation 
showed quantitatively good agreement, 
guaranteeing the validities of the formula. 


The author would like to acknowledge 
H. Imamura, T. Yorozu, H. Kubota, H. Maehara, and S. Yuasa 
for their valuable discussions. 
This work was supported by JSPS KAKENHI Grant-in-Aid for Young Scientists (B) 25790044. 


\appendix

\section{Transformation to rotating frame}

The transformation from the laboratory frame to the rotating frame 
is described by the rotation matrix, 
\begin{equation}
  \mathsf{R}
  =
  \begin{pmatrix}
    \cos(2\pi ft) & \sin(2\pi ft) & 0 \\
    -\sin(2\pi ft) & \cos(2\pi ft) & 0 \\
    0 & 0 & 1
  \end{pmatrix}.
\end{equation}
For example, 
the relation between $\mathbf{m}$ and $\mathbf{m}^{\prime}$ is given by 
$\mathbf{m}^{\prime}=\mathsf{R}\mathbf{m}$. 
Similarly, the field $\mathbf{H}$ is transformed as 
$\mathbf{H}^{\prime}=\mathsf{R}\mathbf{H}
  = H_{\rm ac} \mathbf{e}_{x^{\prime}} + (-H + H_{\rm K}m_{z^{\prime}}) \mathbf{e}_{z^{\prime}}$, 
where $\mathbf{H}^{\prime}$ relates to 
$\bm{\mathcal{B}}$ in Eq. (\ref{eq:LLG}) as 
$\bm{\mathcal{B}}=\mathbf{H}^{\prime}-(2\pi f/\gamma) \mathbf{e}_{z^{\prime}}$. 
Also, $d \mathbf{m}/dt$ should be replaced by $d \mathbf{m}^{\prime}/dt - 2\pi f \mathbf{m}^{\prime} \times \mathbf{e}_{z^{\prime}}$. 
Then, Eq. (\ref{eq:LLG_orig}) is transformed as 
\begin{equation}
\begin{split}
  \frac{d \mathbf{m}^{\prime}}{dt}
  =&
  -\gamma
  \mathbf{m}^{\prime}
  \times
  \left(
    \mathbf{H}^{\prime}
    -
    \frac{2\pi f}{\gamma}
    \mathbf{e}_{z^{\prime}}
  \right)
\\
  &-
  \alpha
  \gamma
  \mathbf{m}^{\prime}
  \times
  \left(
    \mathbf{m}^{\prime}
    \times
    \mathbf{H}^{\prime}
  \right).
  \label{eq:LLG_rotate}
\end{split}
\end{equation}
Equation (\ref{eq:LLG_rotate}) is equivalent to Eq. (\ref{eq:LLG}). 
However, for the following reason, 
we use Eq. (\ref{eq:LLG}) instead of Eq. (\ref{eq:LLG_rotate}). 
As shown in Secs. \ref{sec:Reversal in low frequency region} and \ref{sec:Reversal in high frequency region}, 
a potential map is useful to investigate the magnetization dynamics. 
The potential is usually defined as the integral of the field 
with respect to the magnetization \cite{landau80}, 
where the field appears in both the conservative and the damping torques of the LLG equation. 
When we use Eq. (\ref{eq:LLG_rotate}), 
the definition of the potential is not clear 
because the fields that appeared in the conservative torque, $\mathbf{H}^{\prime}-(2\pi f/\gamma) \mathbf{e}_{z^{\prime}}$, 
and in the damping torque, $\mathbf{H}^{\prime}$, are different. 
On the other hand, when we use Eq. (\ref{eq:LLG}), 
the potential can be well defined as $\mathscr{E}= - M \int d \mathbf{m}\cdot \bm{\mathcal{B}}$. 
Therefore, we express the LLG equation in the rotating frame 
in the form of Eq. (\ref{eq:LLG}). 


\section{Calculation procedures of Eqs. (\ref{eq:W_s}) and (\ref{eq:W_alpha})}
\label{sec:Appendix_B}

Equations (\ref{eq:W_s}) and (\ref{eq:W_alpha}) can be calculated 
without the time-dependent solution of $\mathbf{m}^{\prime}(t)$ obtained from Eq. (\ref{eq:LLG}). 
Using the conservative torque term of the LLG equation, 
the integration variable can be transformed from the time $t$ to $m_{z^{\prime}}$, 
i.e., 
from $\oint dt$ to $2 \int d m_{z^{\prime}}/(\gamma H_{\rm ac}m_{y^{\prime}})$, 
where the numerical factor $2$ appears 
by restricting the integral range to $m_{y^{\prime}}>0$ and 
due to the symmetry of the system with respect to the $x^{\prime}z^{\prime}$ plane. 
Because the LLG equation conserves the magnetization magnitude, 
$m_{y^{\prime}}$ appearing in Eqs. (\ref{eq:W_s}) and (\ref{eq:W_alpha}) 
can be replaced by 
$\sqrt{1-m_{x^{\prime}}^{2}-m_{z^{\prime}}^{2}}$. 
Also, from Eq. (\ref{eq:potential}), 
$m_{x^{\prime}}$ can be expressed in terms of $m_{z^{\prime}}$ as 
\begin{equation}
  m_{x^{\prime}}
  =
  \frac{1}{H_{\rm ac}}
  \left[
    -\frac{\mathscr{E}}{M}
    +
    \left(
      H
      +
      \frac{2\pi f}{\gamma}
    \right)
    m_{z^{\prime}}
    -
    \frac{H_{\rm K}}{2}
    m_{z^{\prime}}
  \right].
\end{equation}
Therefore, the integrands in Eqs. (\ref{eq:W_s}) and (\ref{eq:W_alpha}) can be 
expressed in terms of $m_{z^{\prime}}$ only. 
The integral range can be determined 
from Eq. (\ref{eq:potential}) by fixing the value of $\mathscr{E}$. 



\section{Formulae of reversal field in the case of tensor damping}
\label{sec:Appendix_C}

When the damping constant $\alpha$ is replaced by the tensor damping, 
$\mathscr{W}_{\rm s}$ and $\mathscr{W}_{\alpha}$ in Eqs. (\ref{eq:W_s}) and (\ref{eq:W_alpha}) 
should be redefined as 
\begin{equation}
  \mathscr{W}_{\rm s}
  =
  -2\pi f M 
  \oint dt 
  \bm{\mathcal{B}}
  \cdot
  \overset{\leftrightarrow}{\bm{\alpha}}
  \cdot
  \left[
    \mathbf{e}_{z^{\prime}}
    -
    \left(
      \mathbf{m}^{\prime}
      \cdot
      \mathbf{e}_{z^{\prime}}
    \right)
    \mathbf{m}^{\prime}
  \right],
  \label{eq:W_s_general}
\end{equation}
\begin{equation}
  \mathscr{W}_{\alpha}
  =
  -\gamma M 
  \oint dt 
  \bm{\mathcal{B}}
  \cdot
  \overset{\leftrightarrow}{\bm{\alpha}}
  \cdot
  \left[
    \bm{\mathcal{B}}
    -
    \left(
      \mathbf{m}^{\prime}
      \cdot
      \bm{\mathcal{B}}
    \right)
    \mathbf{m}^{\prime}
  \right],
  \label{eq:W_alpha_general}
\end{equation}
where $\overset{\leftrightarrow}{\bm{\alpha}}$ is the tensor damping in the rotating frame, 
and has nine components $\alpha_{k\ell}$ ($k,\ell=x^{\prime},y^{\prime},z^{\prime}$), in general.
The tensor product is defined as, for example, 
$\bm{\mathcal{B}}\cdot\overset{\leftrightarrow}{\bm{\alpha}}\cdot\bm{\mathcal{B}}=\mathcal{B}_{k}\alpha_{k\ell}\mathcal{B}_{\ell}$. 
According to the discussions in Secs. \ref{sec:Reversal in low frequency region} and \ref{sec:Reversal in high frequency region}, 
and using Eqs. (\ref{eq:W_s_general}) and (\ref{eq:W_alpha_general}), 
two conclusions are obtained for the case of the tensor damping. 
First, Eqs. (\ref{eq:condition_low}) and (\ref{eq:condition_high}) are still applicable to 
estimate the reversal fields in the low- and high-frequency regions, respectively, 
because the explicit forms of $\mathscr{W}_{\rm s}$ and $\mathscr{W}_{\alpha}$ do not affect 
the derivation of these equations. 
Equation (\ref{eq:condition_boundary}) is also applicable to 
determine the boundary between the low- and high-frequency regions. 
However, the argument that Eqs. (\ref{eq:condition_boundary}) and (\ref{eq:condition_high}) are independent of the damping constant 
does not necessarily hold. 
This is because the integrals of Eqs. (\ref{eq:W_s}) and (\ref{eq:W_alpha}) are independent of the scalar damping $\alpha$, 
and therefore, the frequency or field satisfying Eqs. (\ref{eq:condition_boundary}) or (\ref{eq:condition_high}) is also independent of $\alpha$, 
while in the case of the tensor damping, the integrals of Eqs. (\ref{eq:W_s_general}) and (\ref{eq:W_alpha_general}) depend on 
the components of $\overset{\leftrightarrow}{\bm{\alpha}}$, in general. 
Second, Eq. (\ref{eq:condition_low_zero_damping}) is also valid 
because this equation is derived by the condition $\mathscr{E}_{\rm initial}-\mathscr{E}_{\rm saddle}=0$, 
which is independent of the damping. 





\end{document}